\documentclass[10pt,conference]{IEEEtran}
\usepackage[T1]{fontenc}
\usepackage[utf8]{inputenc}
\usepackage{graphicx}
\usepackage{booktabs}
\usepackage{pgfplots}
\pgfplotsset{compat=1.18}
\usepackage{xspace}
\usepackage{amsmath}
\usepackage[hidelinks]{hyperref}
\usepackage{orcidlink}
\usepackage[capitalise]{cleveref}
\newcommand{\nPop}{361,296\xspace}  
\newcommand{\nPopRepos}{25,402\xspace}  
\newcommand{\nAllPRs}{7,685,281\xspace}  
\newcommand{\nCand}{108,708\xspace}  
\newcommand{\pctCand}{30.1\%\xspace}  
\newcommand{\nPilot}{2,000\xspace}  
\newcommand{\pctPilotRuntime}{23.3\%\xspace}  
\newcommand{\pctPilotNone}{68.9\%\xspace}  
\newcommand{\nStratClassified}{5,900\xspace}  
\newcommand{\nRuntimePRs}{1,343\xspace}  
\newcommand{\haikuKappa}{0.79\xspace}  
\newcommand{\haikuAgreeN}{66\xspace}  
\newcommand{\haikuAgree}{91\%\xspace}  
\newcommand{\nClaims}{1,343\xspace}  
\newcommand{\nNoMech}{189\xspace}  
\newcommand{\pctNoMech}{14.1\%\xspace}  
\newcommand{\nEffect}{422\xspace}  
\newcommand{\pctEffect}{31.4\%\xspace}  
\newcommand{\evBenchmark}{221\xspace}  
\newcommand{\evTest}{347\xspace}  
\newcommand{\evReasoning}{500\xspace}  
\newcommand{\evNone}{269\xspace}  
\newcommand{\evProfile}{6\xspace}  
\newcommand{\pctReasonNone}{57.3\%\xspace}  
\newcommand{\pctBenchmark}{16.5\%\xspace}  
\newcommand{\mechAlgorithmic}{9.5\%\xspace}  
\newcommand{\mechAsyncIo}{5.5\%\xspace}  
\newcommand{\mechAvoidRepeatedIo}{2.6\%\xspace}  
\newcommand{\mechCacheResults}{15.3\%\xspace}  
\newcommand{\mechConcurrencyTuning}{4.2\%\xspace}  
\newcommand{\mechFasterDbQuery}{3.7\%\xspace}  
\newcommand{\mechFasterLibrary}{4.2\%\xspace}  
\newcommand{\mechFewerDbQueries}{3.6\%\xspace}  
\newcommand{\mechLazyLoading}{4.1\%\xspace}  
\newcommand{\mechLessLocking}{1.7\%\xspace}  
\newcommand{\mechLessMemory}{16.6\%\xspace}  
\newcommand{\mechParallelize}{6.9\%\xspace}  
\newcommand{\mechPrecompute}{5.4\%\xspace}  
\newcommand{\mechReuseResources}{3.9\%\xspace}  
\newcommand{\mechSkipUnneededWork}{20.6\%\xspace}  
\newcommand{\mechVectorizeLocality}{3.3\%\xspace}  
\newcommand{\nMeasurableClaims}{1,030\xspace}  
\newcommand{\nGhError}{19\xspace}  
\newcommand{\nNoPyFinal}{62\xspace}  
\newcommand{\nJobs}{949\xspace}  
\newcommand{\nVMs}{3\xspace}  
\newcommand{\nControls}{12\xspace}  
\newcommand{\nChecks}{22\xspace}  
\newcommand{\ctrlPassMin}{21\xspace}  
\newcommand{\ctrlPassMax}{22\xspace}  
\newcommand{\counterErrMax}{3.5\times10^{-5}\xspace}  
\newcommand{\faultRatioMed}{0.943\xspace}  

\newcommand{\nMeasured}{91\xspace}  
\newcommand{\nExcludedReview}{4\xspace}  
\newcommand{\nExcludedNoisy}{1\xspace}  
\newcommand{\nAnalysed}{82\xspace}  
\newcommand{\nReposAnalysed}{53\xspace}  
\newcommand{\maxPerRepo}{15\xspace}  
\newcommand{\nFaster}{32\xspace}  
\newcommand{\nSlower}{22\xspace}  
\newcommand{\nNoChange}{28\xspace}  
\newcommand{\nHeld}{34\xspace}  
\newcommand{\qFH}{28\xspace}  
\newcommand{\qFN}{4\xspace}  
\newcommand{\qNH}{3\xspace}  
\newcommand{\qNN}{25\xspace}  
\newcommand{\qSH}{3\xspace}  
\newcommand{\qSN}{19\xspace}  
\newcommand{\nBehaviour}{42\xspace}  
\newcommand{\nHiddenCost}{21\xspace}  
\newcommand{\aaMin}{0.965\xspace}  
\newcommand{\aaMax}{1.009\xspace}  
\newcommand{\nLowPy}{4\xspace}  
\newcommand{\nAttempted}{150\xspace}  
\newcommand{\funChangeUnreachable}{2\xspace}  
\newcommand{\funFailedCheck}{16\xspace}  
\newcommand{\funInstallFailed}{2\xspace}  
\newcommand{\funNeedsExternalService}{13\xspace}  
\newcommand{\funNeedsGpu}{4\xspace}  
\newcommand{\funNewCodeOnly}{14\xspace}  
\newcommand{\funNotPythonChange}{6\xspace}  
\newcommand{\funOther}{2\xspace}  
\newcommand{\funPassed}{91\xspace}  
\newcommand{\pctPassed}{61\%\xspace}  
\newcommand{\usdPerBuild}{\$0.37\xspace}  

\newcommand{\caseAWall}{0.88$\times$}
\newcommand{\caseAWallBase}{2.60}
\newcommand{\caseAWallHead}{2.28}
\newcommand{\caseASys}{1.41$\times$}

\newcommand{\caseASyscalls}{3.49$\times$}
\newcommand{\caseASyscallsBase}{74,797}
\newcommand{\caseASyscallsHead}{261,258}
\newcommand{\caseAPread}{11.56$\times$}

\newcommand{\caseABytes}{2.47$\times$}

\newcommand{\caseATask}{1.32$\times$}

\newcommand{\caseAThreadsBase}{19}
\newcommand{\caseAThreadsHead}{32}
\newcommand{\caseBWall}{11.35$\times$}
\newcommand{\caseBWallBase}{0.90}
\newcommand{\caseBWallHead}{10.19}
\newcommand{\caseBGetdents}{41.82$\times$}
\newcommand{\caseBGetdentsBase}{294}
\newcommand{\caseBGetdentsHead}{12,294}
\newcommand{\caseBInstr}{17.88$\times$}

\newcommand{\caseCWall}{0.82$\times$}

\newcommand{\caseCRead}{0.24$\times$}
\newcommand{\caseCReadBase}{17,215}
\newcommand{\caseCReadHead}{4,063}
\newcommand{\caseCInstr}{0.73$\times$}

\newcommand{\caseDWall}{1.47$\times$}
\newcommand{\caseDWallBase}{4.32}
\newcommand{\caseDWallHead}{6.32}
\newcommand{\caseDOffcpu}{0.87$\times$}

\newcommand{\caseDSyscalls}{2.16$\times$}

\newcommand{\caseDFutexc}{2.86$\times$}

\newcommand{\caseDTask}{1.43$\times$}

\newcommand{\caseDSocketHead}{6}

\newcommand{\caseDConnectHead}{4}

\newcommand{\aaSigTests}{1,796\xspace}

\newcommand{\aaSigRate}{0.0\%\xspace}

\newcommand{\nSame}{40\xspace}
\newcommand{\qFHSame}{12\xspace}
\newcommand{\qFNSame}{2\xspace}

\newcommand{\nChg}{42\xspace}
\newcommand{\qFHChg}{16\xspace}
\newcommand{\qFNChg}{2\xspace}

\newcommand{\kClaimedFast}{48.8\%\xspace}
\newcommand{\kUnclaimedFast}{28.7\%\xspace}
\newcommand{\kClaimedSlow}{11.1\%\xspace}
\newcommand{\kUnclaimedSlow}{3.6\%\xspace}
\newcommand{\iClaimedFast}{87.9\%\xspace}
\newcommand{\iUnclaimedFast}{76.7\%\xspace}
\newcommand{\kRatioFast}{1.7\xspace}
\newcommand{\kRatioSlow}{3.1\xspace}
\newcommand{\iRatioFast}{1.15\xspace}
\newcommand{\nHeldKernel}{20\xspace}
\newcommand{\nHeldFastKernel}{14\xspace}
\newcommand{\nHeldFast}{28\xspace}
\newcommand{\nHeldHwOnly}{4\xspace}

\newcommand{\sys}{\textsc{Below}\xspace}

\begin{document}
\bstctlcite{IEEEexample:BSTcontrol}
\title{Below the Stopwatch: Checking the Performance Claims of Coding Agents in the Kernel}

\author{
\IEEEauthorblockN{Yuvraj Sehgal\,\orcidlink{0009-0002-1980-604X}}
\IEEEauthorblockA{\textit{Department of Computer Science}\\
\textit{Brock University}\\
St.\ Catharines, Ontario, Canada\\
ys19rk@brocku.ca}
\and[\hspace{0.6in}]
\IEEEauthorblockN{Sneh Patel\,\orcidlink{0000-0002-1985-8715}}
\IEEEauthorblockA{\textit{Department of Computer Science}\\
\textit{Brock University}\\
St.\ Catharines, Ontario, Canada\\
sp18oo@brocku.ca}
\and[\hfill\vspace{1em}\break\hspace*{\fill}]
\IEEEauthorblockN{Mahsa Panahandeh\,\orcidlink{0000-0002-6369-8982}}
\IEEEauthorblockA{\textit{School of Electrical Engineering}\\
\textit{and Computer Science}\\
\textit{University of Ottawa}\\
Ottawa, Ontario, Canada\\
mpanahan@uottawa.ca}
\and[\hspace{0.6in}]
\IEEEauthorblockN{Naser Ezzati-Jivan\,\orcidlink{0000-0003-1435-6297}}
\IEEEauthorblockA{\textit{Department of Computer Science}\\
\textit{Brock University}\\
St.\ Catharines, Ontario, Canada\\
nezzatijivan@brocku.ca}
}

\maketitle

\begin{abstract}
Coding agents open many pull requests (PRs) that claim to make software
faster. Most such claims also give a reason, such as ``cache the result''
or ``read the file once''. A stopwatch can show that code got faster, but
not that it got faster for the stated reason. We present \sys, a
pipeline that checks these reasons at the kernel level. From \nRuntimePRs
runtime-performance PRs in the AIDev dataset, \sys extracts the mechanism each PR claims and maps it to the kernel
and hardware signals that must change if the claim is true. A build agent
that never sees the claim rebuilds the code before and after each PR.
\sys then measures both versions with plain timing runs, hardware counters and
LTTng kernel traces. Most claims are not measured: \pctReasonNone of the
PRs offer only reasoning or no evidence, and only \pctEffect state an
effect size. Of \nAnalysed analyzed PRs, \nFaster ran faster, \nNoChange
showed no clear change and \nSlower ran slower. The expected signal of the
claimed mechanism appeared in \nHeld PRs (\nHeldKernel in kernel-trace
signals alone), and \nBehaviour PRs also changed
their output. Kernel signals move about \kRatioFast times as often for a
faster PR's claimed mechanism as for mechanisms it did not claim; the
instruction count barely separates them (\iRatioFast times).

\end{abstract}

\section{Introduction}
\label{sec:intro}

Coding agents now open pull requests (PRs) at scale. The latest release of
the AIDev dataset~\cite{li2026aidev} alone records \nAllPRs PRs written by
six agents~\cite{aidev2026v5}. Many of these PRs claim to make software
faster or lighter. Existing studies of such performance PRs read their
text~\cite{opu2025bertopic,peng2025optimization,
dovhoshliubnyi2026mutation,peng2026validation}, or re-execute a few dozen
of them~\cite{qi2026merged}. They find that agents rely more on reasoning
and less on benchmarks than humans do~\cite{peng2026validation}, and that
re-executed claims often do not hold~\cite{qi2026merged}.

A stopwatch tells us whether a change is faster, but not \emph{why}. Yet a
performance PR usually names a mechanism: \emph{cache the result},
\emph{read the file once}, \emph{stream in 1\,MB chunks}, \emph{remove the
lock}. Each mechanism should leave a trace in the operating system, which
we call its \emph{footprint}: fewer \texttt{read} system calls, fewer page
faults, or less time waiting on a lock (a \texttt{futex} wait). Two cases
matter. If the code is faster but the footprint is missing, the stated
reason does not explain the speed-up. If the footprint is present but the
code is not faster, the mechanism was not the bottleneck. A stopwatch sees
neither case.

We present \sys, a method and pipeline that checks agents' performance
claims at the kernel level. Its name comes from its main idea: it looks
below the stopwatch, at what the operating system and the hardware
actually did. \sys (i)~finds performance PRs and extracts the
mechanism each one claims; (ii)~maps each mechanism to the kernel and
hardware signals that must change if the claim is true; (iii)~uses a
\emph{blind} build agent, which never sees the claim, to rebuild the code
before and after each PR and to write a workload that runs the changed
code; and (iv)~measures both versions with plain timing runs, hardware
counters and full LTTng kernel traces~\cite{desnoyers2006lttng}. We ask
three research questions.
\textbf{RQ1:} What mechanisms do agents claim, and how do they support
them? \textbf{RQ2:} Can a blind build agent write workloads that run the
changed code? \textbf{RQ3:} Do claimed mechanisms leave their footprint,
and do kernel signals tell mechanisms apart better than timing and
instruction counts? We answer them for \nClaims performance claims and
\nAnalysed analyzed PRs.

\section{Approach}
\label{sec:approach}

\begin{figure*}[t]
\centering
\includegraphics[width=\textwidth]{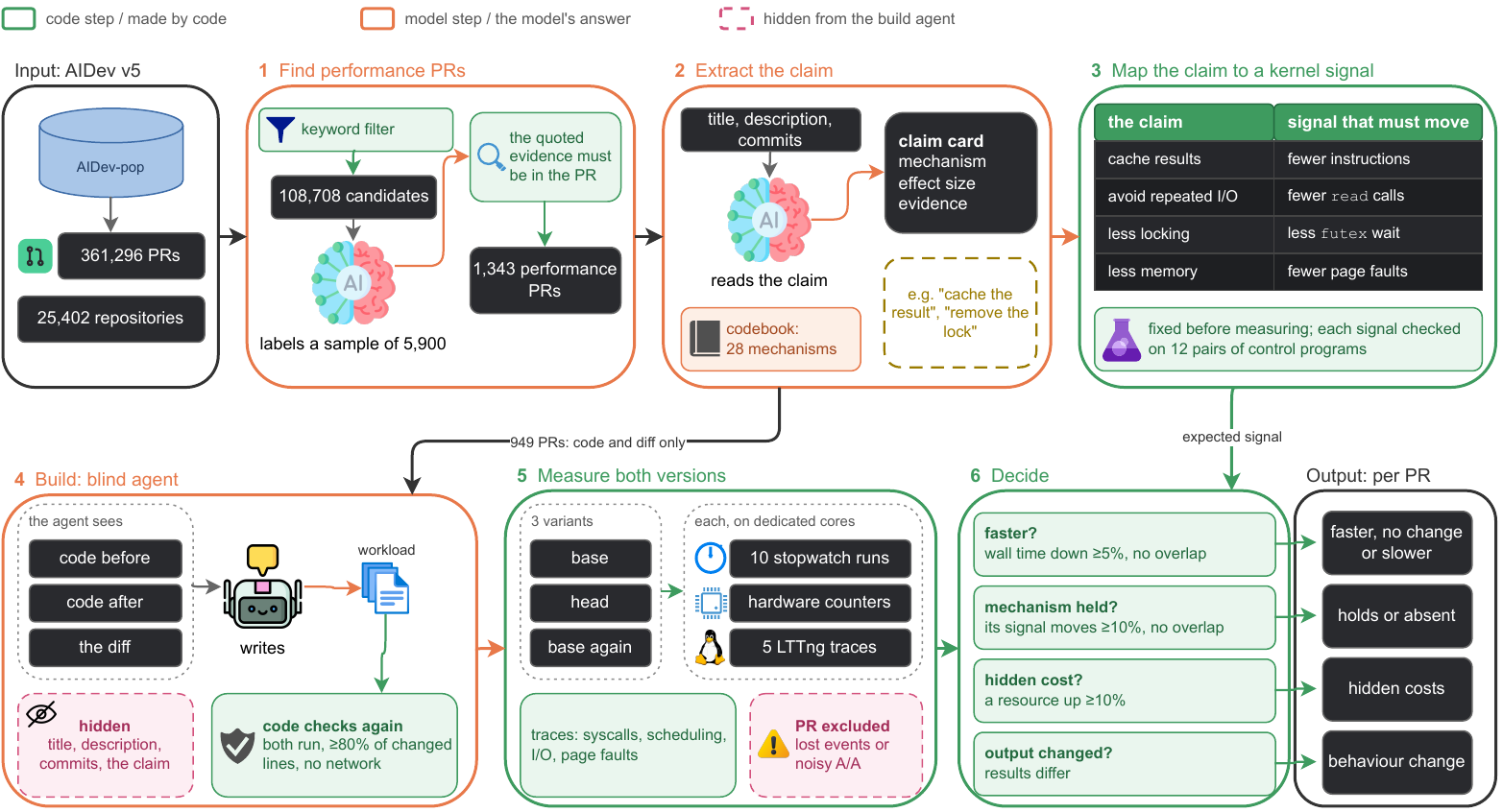}
\caption{Overview of \sys. Orange boxes are model steps and green boxes
are code. The build agent (step~4) sees only the code and the diff; the
PR text and the claim are hidden from it (pink). The expected signal from
step~3 goes straight to the decision in step~6.}
\label{fig:overview}
\end{figure*}

Figure~\ref{fig:overview} shows the six steps of the pipeline. Throughout,
\emph{base} is the code just before a PR and \emph{head} is the code with
the PR applied.

\subsection{Performance PRs and their claims}
\label{sec:data}
We use version~5 of AIDev~\cite{li2026aidev,aidev2026v5}, pinned to one
dataset commit. We keep its \emph{AIDev-pop} subset: \nPop agent PRs from
\nPopRepos repositories with more than 100 stars. A keyword filter over
titles, descriptions and commit messages keeps \nCand candidates
(\pctCand); it first removes template text that some agents add to every
PR. A language model then labels each candidate, following written
labelling rules (a codebook): is the PR's goal to make the software run
faster or use fewer resources? On a separate pilot sample of \nPilot
candidates, \pctPilotRuntime are such runtime-performance PRs and
\pctPilotNone are not about performance at all.

We study PRs that change Python code. For each agent, we label its
candidates in a fixed random order until we reach a target number of
runtime-performance PRs; for the three agents with few PRs, we label all
of them. This labels \nStratClassified PRs and yields \nRuntimePRs
runtime-performance PRs. The labels come from Claude models. On \haikuAgreeN
PRs labelled by two different models, the labels agree on \haikuAgree (Cohen's
$\kappa=\haikuKappa$). Every label must quote its evidence word for word
from the PR, and code checks that the quote is there.

A second codebook extracts, from each runtime-performance PR, the
mechanisms it claims (28 categories), the effect size it claims, and the
evidence it offers. We built its categories by reading 100 PRs, revised
it once, and then froze it. Table~\ref{tab:oracle} lists the measurable
mechanisms.

\subsection{From claimed mechanism to kernel signal}
\label{sec:oracle}
Before any measurement, we mapped each mechanism to the signals that must
change if the claim is true (Table~\ref{tab:oracle}). A mechanism
\emph{holds} if at least one of its signals changes as expected. To check
that each signal responds as designed, we wrote twelve pairs of small
control programs, one pair per signal family. For example, one program
reads a file 2,000 times and its partner reads it once; another has two
threads share one lock, and its partner gives each thread its own counter.

\begin{table}[t]
\caption{Claimed mechanism, share of the \nClaims runtime-performance PRs
that claim it (a PR may claim several), and the signals that must move (one is enough).}
\label{tab:oracle}
\centering\scriptsize
\setlength{\tabcolsep}{3pt}
\begin{tabular}{@{}lrl@{}}
\toprule
mechanism claimed & PRs & signal (head vs.\ base) \\
\midrule
skip unneeded work & \mechSkipUnneededWork & fewer syscalls / instructions \\
less memory, fewer copies & \mechLessMemory & fewer page faults, page allocations \\
cache results & \mechCacheResults & fewer instructions or syscalls \\
algorithmic & \mechAlgorithmic & fewer instructions \\
parallelize & \mechParallelize & wall time down, CPU time not down \\
async / non-blocking I/O & \mechAsyncIo & less off-CPU time on the main thread \\
precompute & \mechPrecompute & fewer instructions \\
faster library & \mechFasterLibrary & fewer instructions \\
concurrency tuning & \mechConcurrencyTuning & threads, context switches, futex \\
lazy loading & \mechLazyLoading & fewer \texttt{openat} / bytes read \\
reuse resources & \mechReuseResources & fewer \texttt{socket}/\texttt{connect}, threads, forks \\
DB: faster / fewer queries & \mechFasterDbQuery\,/\,\mechFewerDbQueries & fewer page reads, round trips \\
vectorize, locality & \mechVectorizeLocality & fewer instructions, cache misses \\
avoid repeated I/O & \mechAvoidRepeatedIo & fewer \texttt{openat}/\texttt{read}, bytes \\
less locking & \mechLessLocking & less futex wait \\
\bottomrule
\end{tabular}
\end{table}

\subsection{Blind build agent and measurement}
\label{sec:harness}
For each PR, the harness checks out base and head. Base is the merge base
of the PR, because the base commit that the GitHub API reports can be
later than the branch point and would mix in unrelated changes. The
harness installs both versions with dependencies frozen at the PR's
creation date. It then hands them to a build agent (Claude Code),
which may repair a failed install, the same way on both sides.
The agent is \emph{blind to the claim}. It sees the code, the diff and
the changed files. It does not see the PR title, description, commit
messages or our extracted claim, and the git history is moved out of its
reach. It must write a deterministic workload that runs the changed code
on both versions. It may replace external services with stand-ins, but it
must not patch the code under test. The harness then checks the workload
independently: both versions must run; the workload must run at least
80\% of the executable lines the PR adds or changes (50\% for PRs with
more than 200 such lines); base must take 0.5--30\,s; and the workload must
run without a network. It also compares a hash of each version's output.
Finally, we audit the agent's transcript for any access to the PR on
GitHub.

Measurements run on \nVMs identical Google Cloud C4 virtual machines with
the performance-monitoring unit enabled, one job per machine at a time.
Workloads run on dedicated CPU cores (one, or two for parallel
workloads). We confine all system services to the other cores with
systemd. We do not use the kernel's \texttt{isolcpus} option, because it
disables load balancing: in a test, a two-process pool ran on a single
core. Cloud machines make performance results vary~\cite{laaber2019cloud},
small setup details can bias a comparison~\cite{mytkowicz2009wrong}, and
versions should be compared over repeated runs~\cite{georges2007rigorous}.
We therefore measure three variants per PR: base, head, and base again.
The second base run is an \emph{A/A control}: base against itself shows
how much the machine alone varies. After 3 warm-up runs each of base and
head, we interleave the three variants in every round, so slow drift
affects all three alike. Each variant gets 10 timing runs without
tracing, so the tracer cannot slow them. It also gets 5 runs of each of 3
hardware-counter groups, with at most four programmable counters per
group, so that no counter is sampled part-time (multiplexed). Finally,
it gets 5 full LTTng kernel traces: all system calls, scheduling, block
I/O, interrupts, page faults and page allocations. We keep only the
events of the workload's own processes and attribute file I/O to each
file. PRs with any lost trace event or multiplexed counter are excluded.

The control programs confirm the setup. On each machine, the
\nControls pairs pass \ctrlPassMin--\ctrlPassMax of \nChecks checks. The
only failure is the number of context switches under lock contention,
which varies from run to run; the main signal for locking, \texttt{futex}
wait time, passes. A loop with a known number of instructions is counted
within a relative error of $\counterErrMax$. LTTng's page-fault events
and perf's page-fault counter agree at a median ratio of \faultRatioMed.

\section{Results}
\label{sec:results}

\subsection{RQ1: What agents claim, and how they back it}
\label{sec:claims}
Of the \nClaims runtime-performance PRs, \nNoMech (\pctNoMech) name no
mechanism at all. The most common mechanisms are skipping unneeded work
(\mechSkipUnneededWork), reducing memory use or copies (\mechLessMemory)
and caching results (\mechCacheResults) (Table~\ref{tab:oracle}). Most
claims are not measured: \evReasoning PRs support their claim by reasoning
only and \evNone offer no evidence (together \pctReasonNone), while
\evBenchmark cite a benchmark (\pctBenchmark), \evTest a test, and
\evProfile a profile.
Only \nEffect (\pctEffect) state an effect size. In total, \nMeasurableClaims
PRs claim at least one mechanism that has a kernel or hardware signal. The
rest claim only GPU, external-API, user-interface or uncategorized
mechanisms, or no mechanism.

\subsection{RQ2: Building the PRs}
\label{sec:funnel}
Of the \nMeasurableClaims PRs, \nGhError are no longer available on GitHub.
Another \nNoPyFinal do not change Python code in their final diff; AIDev's
per-commit file lists also include commits merged in from the base
branch. This leaves \nJobs build jobs, which we process in one fixed
random order. In the first batch, the first \nAttempted of that order,
the build agent wrote a workload that passed all harness checks for
\funPassed PRs (\pctPassed), at about \usdPerBuild per PR. So for most
PRs, a blind agent can build a usable workload. The others need an
external service (\funNeedsExternalService), add only new code that base
cannot run (\funNewCodeOnly), need a GPU (\funNeedsGpu), failed to install
(\funInstallFailed), had unreachable changed code (\funChangeUnreachable),
or did not change Python in a runnable way (\funNotPythonChange). For
\funFailedCheck PRs the agent reported success but the independent check
failed, and \funOther could not be classified.

\subsection{RQ3: Do the claims hold below the stopwatch?}
\label{sec:measure}
We analyze \nAnalysed of the \nMeasured measured PRs, from \nReposAnalysed repositories. Each
claims at least one checkable mechanism and changes mostly Python code.
We set aside \nLowPy PRs whose change is mostly non-Python code, because
their claim may concern code we do not run. The manual review excludes
\nExcludedReview PRs: in three, the claimed mechanism lies in code the
workload never runs; in one, base fails on the measured path. We exclude
\nExcludedNoisy PR whose A/A control (base against base) itself differs by
more than 5\%. For all other PRs, the A/A wall-time ratio stays within
\aaMin--\aaMax, so the machine's own noise is small. A PR is
\emph{faster} if its median wall time drops by at least 5\% and every head
run is faster than every base run. A signal \emph{moves} if it changes by
at least 10\% in the claimed direction and every head run lies beyond
every base run.

\begin{table}[t]
\caption{First measurements: wall time against the claimed mechanism, for
the \nAnalysed analyzed PRs.}
\label{tab:quadrant}
\centering\footnotesize
\begin{tabular}{@{}lccc@{}}
\toprule
 & faster & no clear change & slower \\
\midrule
claimed mechanism holds & \qFH & \qNH & \qSH \\
claimed mechanism absent & \qFN & \qNN & \qSN \\
\bottomrule
\end{tabular}
\end{table}

\begin{figure}[t]
\centering
\begin{tikzpicture}
\begin{semilogxaxis}[width=\columnwidth, height=4.2cm,
  xlabel={wall time, head / base (log scale)}, ylabel={PR (sorted)},
  ytick=\empty, xmin=0.03, xmax=15, enlarge y limits=0.07,
  xtick={0.05,0.1,0.2,0.5,1,2,5,10}, xticklabels={0.05,0.1,0.2,0.5,1,2,5,10},
  legend style={font=\scriptsize, at={(0.98,0.04)}, anchor=south east},
  label style={font=\footnotesize}, tick label style={font=\footnotesize}]
\fill[gray!20] ({axis cs:0.95,1}|-{rel axis cs:0,0}) rectangle ({axis cs:1.05,1}|-{rel axis cs:0,1});
\draw[gray] ({axis cs:1,1}|-{rel axis cs:0,0}) -- ({axis cs:1,1}|-{rel axis cs:0,1});
\addplot[only marks, mark=*, black] table[x=ratio, y=rank,
  restrict expr to domain={\thisrow{held}}{1:1}] {fig_wall.dat};
\addplot[only marks, mark=o, black] table[x=ratio, y=rank,
  restrict expr to domain={\thisrow{held}}{0:0}] {fig_wall.dat};
\addplot[only marks, mark=none, point meta=explicit,
  nodes near coords={\pgfmathprintnumber[fixed,precision=2]{\pgfplotspointmeta}$\times$},
  nodes near coords align={anchor=east},
  every node near coord/.append style={font=\scriptsize, xshift=-2pt}]
  table[x=ratio, y=rank, meta=ratio,
  restrict expr to domain={\thisrow{ratio}}{10:1000}] {fig_wall.dat};
\legend{mechanism holds, mechanism absent}
\end{semilogxaxis}
\end{tikzpicture}
\caption{Wall-time ratio of each analyzed PR. The grey band is the
$\pm$5\% zone counted as no clear change.}
\label{fig:wall}
\end{figure}
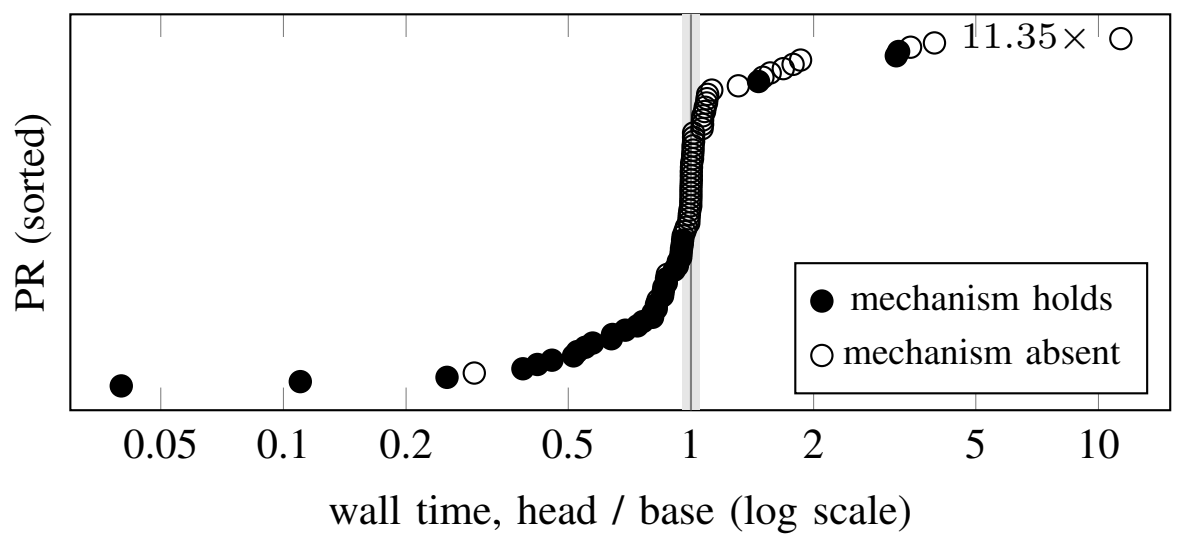

Table~\ref{tab:quadrant} and Figure~\ref{fig:wall} show the results. Of
the \nAnalysed PRs, \nFaster are faster, \nNoChange show no clear change
and \nSlower are slower. The claimed mechanism holds in \nHeld. When a PR
is faster, its claimed mechanism usually holds (\qFH of \nFaster); when it
is slower, it rarely does (\qSH of \nSlower). In this sample, a missing
footprint rarely comes with a speed-up. Many of these holds rest on
hardware counters, mainly the instruction count: with kernel-trace signals
alone, the mechanism holds in \nHeldKernel of the \nHeld PRs, and in
\nHeldFastKernel of the \nHeldFast faster ones (\nHeldHwOnly PRs claim only mechanisms that
have no kernel signal). The rule does not fire by chance:
applied to base against base again, no signal moves in any of
\aaSigTests tests (\aaSigRate).

\textbf{Do kernel signals tell mechanisms apart?} A signal is only useful
if it moves for the claimed mechanism more often than for other
mechanisms. We therefore also test every PR against the signals of
mechanisms it did \emph{not} claim. The instruction count barely separates
the two: among faster PRs, it drops for \iClaimedFast of claimed and
\iUnclaimedFast of unclaimed mechanisms (ratio \iRatioFast), because
almost any speed-up removes instructions. Kernel-trace signals alone
(system calls, futex, off-CPU time, page allocations, bytes read,
processes) separate them better: \kClaimedFast against \kUnclaimedFast
for faster PRs (ratio \kRatioFast), and \kClaimedSlow against
\kUnclaimedSlow for the other PRs (ratio \kRatioSlow). So the kernel view
carries information about the mechanism that a stopwatch and an
instruction counter lack. It is not perfectly specific, however: a held
mechanism is evidence consistent with the claim, not proof of it.

In \nBehaviour PRs, the output of the workload also changes, for example
through new defaults or new error messages; Qi et al.\ report the same
for merged agent fixes~\cite{qi2026merged}. Such changes do not drive the
pattern above. Of the \nSame PRs with unchanged output, \qFHSame are
faster with the mechanism present and \qFNSame faster without it; of the
\nChg PRs with changed output, the numbers are \qFHChg and \qFNChg.
Finally, \nHiddenCost PRs make at least one resource clearly worse: system
time, memory, page faults, system calls, instructions, CPU time or bytes
read.

\begin{table}[t]
\caption{Analysed PRs by agent and by the evidence the PR offers for its
claim: number of PRs, faster, and claimed mechanism held.}
\label{tab:agents}
\centering\footnotesize
\begin{tabular}{@{}lrrr@{}}
\toprule
 & PRs & faster & held \\
\midrule
Claude Code & 22 & 9 & 10 \\
Jules & 15 & 6 & 9 \\
Devin & 15 & 3 & 3 \\
Codex & 12 & 4 & 5 \\
Copilot & 10 & 6 & 5 \\
Cursor & 8 & 4 & 2 \\
\midrule
reasoning only & 38 & 12 & 10 \\
test & 21 & 7 & 9 \\
benchmark & 14 & 8 & 10 \\
no evidence & 6 & 5 & 5 \\
profile & 3 & 0 & 0 \\
\bottomrule
\end{tabular}

\end{table}

\textbf{Agents and evidence.} In Table~\ref{tab:agents}, each PR has one
evidence label (types are not ranked). Claims backed by a benchmark were
faster more often than claims backed by reasoning alone. The groups are
small (the 6 PRs with no evidence are too few to compare), so this is a
first observation, not a finding.

Four PRs show what the kernel view adds. All ratios below use unrounded
medians.

\paragraph{Faster, but not for the claimed reason}
iPhotron PR~\#453 (Copilot) claims to remove the slowdown of deep SQL
pagination and to load assets on demand instead of all at once. Head is faster
(base \caseAWallBase\,s, head \caseAWallHead\,s; \caseAWall). Yet the kernel shows more work, not less: \caseASyscalls\
as many system calls (base \caseASyscallsBase, head \caseASyscallsHead),
\caseAPread\ as many \texttt{pread64} calls, \caseABytes\ the bytes read
and \caseASys\ the system time. Total CPU time grows \caseATask, while the
process runs \caseAThreadsHead\ threads instead of \caseAThreadsBase. Peak
memory does not change. A stopwatch would credit the claim; the trace does
not support it.

\paragraph{Slower, and the trace shows where}
iPhotron PR~\#564 (Jules) uses pre-generated thumbnails. Our workload keeps
500 unrelated files in the thumbnail folder. On it, head is \caseBWall\
slower than base (base \caseBWallBase\,s, head \caseBWallHead\,s), and its
output changes. The trace points to a likely cause: calls to
\texttt{getdents64}, the system call that lists a directory, grow
\caseBGetdents\ (base \caseBGetdentsBase, head \caseBGetdentsHead), and
instructions grow \caseBInstr.

\paragraph{Half the mechanism happens, and it hurts}
A Claude Code PR to gemini-notebook-mcp-cli (\#16) claims non-blocking I/O
and HTTP connection pooling. The first part appears: the main thread
spends \caseDOffcpu\ of its former time off the CPU. The second does not:
both versions open \caseDSocketHead\ sockets and \caseDConnectHead\
connections to the workload's local stand-in server; connection reuse
would still show against a local server. Meanwhile, system
calls grow \caseDSyscalls, \texttt{futex} calls \caseDFutexc\ and CPU time
\caseDTask, and head is \caseDWall\ slower (base \caseDWallBase\,s, head
\caseDWallHead\,s).

\paragraph{A claim that holds}
mlflow PR~\#19520 (Copilot) changes artifact transfer to chunked
streaming. The footprint matches the claim: \texttt{read} calls fall to
\caseCRead\ (base \caseCReadBase, head \caseCReadHead) with equal bytes
read. Instructions fall to \caseCInstr\ and wall time to
\caseCWall\ of base.

\section{Discussion, threats and related work}
\label{sec:discussion}

\textbf{What this means for agent builders and reviewers.} A performance
PR should not be merged on its description alone if its mechanism leaves
no kernel footprint, or if its speed-up comes with a changed output. Once
a workload exists, kernel signals are cheap to collect. Agents could be
asked to attach trace evidence for the mechanism they claim; this would
turn many claims that cannot be checked today into claims that can.

\textbf{Threats to validity.} The sample is small and early (\nAnalysed PRs,
up to \maxPerRepo from one repository), so the shares in
Table~\ref{tab:quadrant} are preliminary. Even with the coverage check, an
agent-written workload may not reflect real use. Because the agent is blind to the claim, it may run a different part
of a large PR than the one the claim is about. The review removed such
PRs when it found them, but others may remain; they would bias results
towards ``mechanism absent''. ``Absent'' can also mean that the workload
never reaches the conditions in which a mechanism matters, such as
repeated calls for a cache or large files for streaming. Stand-ins for
external services cannot show network latency or server-side database
work, although they still show the client's connections and calls. Signals
count the whole process tree, including interpreter start-up, which can
hide small effects. Labels, claims and workloads all come from Claude
models, and Claude Code is one of the agents we study. A human check of a
stratified sample of 300 PRs is prepared but not yet done. Each PR is measured once, on one machine.

\textbf{Related work.} Studies of agent performance PRs describe their
topics~\cite{opu2025bertopic}, optimization
patterns~\cite{peng2025optimization}, mutation
patterns~\cite{dovhoshliubnyi2026mutation} and how agents report
validation~\cite{peng2026validation}, all from text and code. Qi et al.\
re-execute 53 fixes from AIDev~\cite{qi2026merged}. They show whether a
fix delivers its claim; kernel signals add why it does or does not. Gong
et al.\ find agent PR descriptions that claim changes the code does not
implement~\cite{gong2026inconsistency}. We check a claim's effect at run
time, not its presence in the diff. Benchmarks such as
PIE~\cite{shypula2024pie}, EffiBench~\cite{huang2024effibench},
SWE-Perf~\cite{he2025sweperf}, GSO~\cite{shetty2025gso},
PerfBench~\cite{garg2025perfbench}, SWE-fficiency~\cite{ma2025swefficiency}
and FormulaCode~\cite{sehgal2026formulacode} measure whether models or
agents can make given code faster; their reliability has been
questioned~\cite{chen2026benchmarks}. Performance bugs in human-written
code have long been studied~\cite{jin2012perfbugs,zaman2012perfbugs}, and
kernel traces have been used to explain where execution time
goes~\cite{giraldeau2016wait,ezzatijivan2017multiscale}. We use kernel
traces to check the mechanism that agent PRs claim.

\textbf{Conclusion.} Most agent performance claims are not measured. When
a PR is faster, its claimed mechanism usually leaves a footprint, but
only \nHeldFastKernel of these \nHeldFast show it in kernel-trace signals;
the rest show it only in hardware counters such as the instruction count.
Kernel-trace signals tell claimed
from unclaimed mechanisms better than the instruction count does. By looking below the stopwatch, \sys shows why a
change is faster, not only whether it is. The full study of all \nJobs build jobs is
under way. Code, data and all kernel traces are
available\footnote{\url{https://doi.org/10.5281/zenodo.24413315}}.

\section*{Acknowledgment}
In line with the workshop's policy, we disclose substantive use of AI.
Claude models classified the PRs and extracted the claims, and a Claude
Code build agent wrote the workloads. Claude also helped write the
analysis code, summarize related work, prepare the figures and draft this
paper. All numbers are generated from the result files by scripts. Each
measured PR was reviewed before inclusion, and the authors take
responsibility for all content.

\bibliographystyle{IEEEtran}
\bibliography{refs}
\end{document}